# Fuel Target Implosion in Ion beam Inertial Confinement Fusion

Shigeo Kawata#

Graduate School of Engineering, Utsunomiya University

Utsunomiya 321-8585, Japan

The numerical results for the fuel target implosion are presented in order to clarify the target physics in ion beam inertial fusion. The numerical analyses are performed for a direct-driven ion beam target. In the paper the following issues are studied: the beam obliquely incidence on the target surface, the plasma effect on the beam-stopping power, the beam particle energy, the beam time duration, the target radius, the beam input energy and the non-uniformity effect on the fuel target performance. In this paper the beam ions are protons.

1. Introduction

In Ion Beam (IB) Inertial Confinement Fusion (ICF), the main research issues include 1) the IB production[1-3] and the IB focusing, 2) the IB transportation[4-9] in an IB-ICF reactor and 3) the target physics, as well as 4) the reactor issuess. The first and second points has been studied[1-9]. In this paper the third point of the fuel target physics is treated theoretically.

This paper presents the numerical analyses of the target implosion in IB ICF. In this paper IB means a proton beam. The numerical results are based on the one-dimensional (1-D)[10] computer code. In this paper the following important issues are focused and presented: 1) the IB oblique incidence to the target surface, 2) the plasma effect of the IB-stopping power, 3) the IB particle energy, 4) the IB time duration, 5) the target radius and 6) the IB input energy on the fuel target performance. The effects of the non-uniformity on the target implosion have been investigated already in another paper by 3-D numerical computations[11]. The results of the non-uniform implosion effect on the target performance are also briefly summarized at the end of this paper.

1. Fuel Target Implosion in IB ICF

---

# The work was done in the author's Master (April, 1978 – March, 1980) and PhD (April. 1980 – October, 1981) courses at Tokyo Institute of Technology under the supervisor Prof. Keishiro Niu. The report was also included in the report on Heavy Ion Inertial Confinement Fusion Reactor System – HIBLIC (Nov., 1985), published by Nagoya University.

In this section, the characteristic phenomena of the IB target implosion are investigated mainly by the 1D fluid numerical analyses.

The target structure employed in the chapter is presented in Fig. 1. The ions deposit their energy in the tamper through the Coulomb-collisional-stopping power mainly. Therefore, the ion beams create the energy deposition Bragg peak in the Al layer in this specific target structure in Fig. 1. If the target has the structure shown in Fig. 1, the DT fuel is accelerated and compressed efficiently. The numerical computation code is described in the references 10-11).

At first, the typical numerical results are presented. The employed parameter values are listed in Table 1. Figures 2 show the stream lines and Fig. 3 show the space profiles of the ion temperature, the mass density and the pressure. Figure 4 shows the time sequences of the peak ion temperature of the DT fuel, the pellet gain $Q$ (= the output fusion energy / the input IB energy $E_b$) and the density-radius product $\rho R$. Figure 5 presents the implosion efficiencies. In Fig. 5 $\eta_{imp}$ is defined by $(E_{TDT}+E_{KDT})/E_b$, $\eta_{KDT}$ is defined by $E_{KDT}/E_b$ and $\eta_{Kin}$ is defined by $(E_{KDT}+E_{KAlin})/E_b$. Here $E_{TDT}$ shows the thermal energy in the DT fuel, $E_{KDT}$ the kinetic energy in the DT fuel and $E_{KAlin}$ the kinetic energy, which is carried by the inward moving Al layer, that is a part of the payload. The implosion efficiency of $\eta_{Kin}$ has a large value before the void closure time ($\tau_v$). Before the void closure time, the DT fuel is compressed gradually. Then the void is closed and the fuel is compressed further more. These processes are presented in Figs. 2 and 6. In Fig. 6 the notation <α> means the preheat ratio and is defined by (the spatially averaged ion temperature of the DT fuel at time ($t$) / the spatially averaged ion temperature estimated by the adiabatic relation from the initial ion pressure and the density at $t$).

2. IB Oblique Incidence on Target Surface

Usually the incident IB has the finite radius and the target is illuminated by the obliquely incident IB particles[10]. A part of the beam ions hits the target surface normally. But some of the beam ions hit the target surface obliquely. As the maximum incident angle $\theta_m$ increases, the Bragg peak becomes wide and the peak position moves outward in the target tamper as shown in Fig. 7. Therefore the material mass, which moves inward and is compressed with the DT fuel, becomes large as the increase in $\theta_m$. This means that the implosion efficiency decreases with the increase in the incident angle. The relation between the implosion efficiency and $\theta_m$ is shown in Fig. 8. The numerical simulations are carried out in the case as shown in Table 2.

On the other hand, the ion energies of IB would be distributed in the range of $\Delta e_b$. It is clear that the broad energy deposition profile has the same effect as that of the IB incident angle (see Fig. 9).

In summary, $M_{in}$ should be taken to be much less than $M_{out}$. Here $M_{in}$ shows the total mass of the material, which moves inward including the DT fuel, and $M_{out}$ the total mass of the material which moves outward.

3. Plasma Effect on IB Stopping Power

During the IB illumination on the target, the target material is partially ionized. Therefore the free electrons, the bound ones and the ions contribute to the IB stopping power. Especially the plasma produced in the ion beam energy deposition layer behaves as not only the individual single particle gas but also the collective particle gas[12]. The stopping power consists of the Coulomb collisional and the plasma wave one. Therefore the total stopping power increases and the stopping range decreases as the increase in the material temperature in the relatively low temperature range, for example, 0 to 300 eV for the typical case as shown in Fig. 10. From the considerations above and in this section 2, it seems to be apparent that the implosion efficiency becomes low, when the plasma effect on the stopping power is switched on.

In order to check the plasma effect on the target implosion, the numerical simulations are performed for two cases, which are the plasma effect-switched-on and -off ones, as shown in Table 3. As described above, the implosion efficiency $\eta_{KDT}$ has a smaller value in the case in which the plasma effect is switched on, than that in the other case. However, it should be noted that $\eta_{Kin}$ and $\eta_{imp}$ have larger values in the former case than those in the latter. This fact can be explained by the following considerations: in the Al layer, the retained thermal energy, which can not contribute to the implosion until the void closure time, is presented by the notation of $E_{TAl}$ in Table 3. The two values in the column of $E_{TAl}$ are quite different with each other. In the case including the plasma effect, the IB input energy is used efficiently. This result can be explained as follows: when the shortening of the IB stopping range appears in the tamper or the energy deposition layer, the narrower region is heated and expands into the target radius size. Therefore, the thermal energy of the A1 layer is converted efficiently to the kinetic energy. This fact leads to the higher implosion efficiency of $\eta_{imp}$ in the case, in which the plasma effect is switched on.

In summary, the IB deposition region should be small in order to realize the efficient conversion from the thermal to the kinetic energy.

4. IB Particle Energy effect on target implosion

From the result in the above section 3, it is important that the IB energy should be deposited in the narrower region. In order to realize it, it seems to be better to take the lower particle energy $e_b$. Because the stopping range strongly depends on the IB particle energy.

In order to investigate the effect of the difference of the beam ion energy on the implosion, the computations of two cases in Table 4 are performed. As described above, $E_{TAL}$ has the smaller value in the case of the lower $e_b$ (=3MeV) than that in the other case $e_b$ (=5MeV). This fact means that the IB input energy deposited in the Al layer is used more efficiently in the case of $e_b$=3MeV. However, the implosion efficiency is small in the low $e_b$ case, compared with that in the other case. This fact comes from the following two points: 1) the IB input energy into the Pb layer is larger in the low $e_b$ case than the other, because in the low $e_b$ case the deposition region is relatively narrow and the temperature increases rapidly in the deposition region. Therefore, the plasma effect affects strongly on the stopping range and the outer region is heated more. The total retained thermal energy $E_T$ in the outer tamper region is large in the low $e_b$ case, compared with the other. For the higher implosion efficiency it is important to covert the input energy $E_b$ to the fuel kinetic energy efficiently. Therefore, $E_T$ should be kept to be small. 2) Another important point is the preheat temperature $T_{DT}$ of the DT fuel. The rapid increase of the pusher temperature leads to the appearance of the strong shock wave and the strong preheat. In the case of $e_b$=3MeV, $T_{DT}$ has a larger value than that in the other. In order to realize the high density of the DT fuel the preheat temperature should be small.

In summary, it should be pointed out that the input energy should be deposited during a relatively long time interval for the high implosion efficiency and the high density.

5. IB Pulse Length

As shown in above section 4, it is important that the input energy is deposited during a relatively long time interval in the tamper. In Tale 5, the comparison between two cases of $\tau_b$=35.0 nsec and 55.0 nsec is presented. As is expected, the preheat temperature is low in the case of the long pulse duration, compared with that in the other case. But the implosion efficiency has a low value for the case of the short pulse duration. It is explained by the relation between the void closure time and the IB pulse length, as follows: For the longer IB pulse length, the input energy at the later stage of the IB time duration cannot be used effectively for the target implosion, because the only energy which is deposited far before the void closure time contributes to the implosion efficiently. While the input energy, which is deposited near or after the void closure time, cannot be converted to the fuel kinetic energy. In summary, it should be noted that the ion beam pulse duration must be much shorter than that of the void closure time.

6. Target Radius

The void closure time is strongly depends on the target radius, because the implosion velocity is nearly same in many usual cases and ~$3 \times 10^7$ cm/sec. Therefore, to realize the relation of $\tau_b <$ (the void closure time $\tau_v$), the large void or the large radius target should be chosen, except the considerations about the problems of the Rayleigh-Taylor (R-T) instability.

In Table 6, the numerical results are summarized for the comparison between the large and small void targets. As presented in Table 6, the implosion efficiencies and the other fusion parameters have the better values for the larger target. At the same time the non-uniformity of the fuel implosion must be studied.

7. IB Input Energy

In Fig. 11, the target gain is plotted for the IB total input energy. In Fig. 11 only the DT total mass is optimized and the other parameters are shown in the figure. The gain curve is not flat and increases with the increase in the input energy. This fact comes from the following: the required minimum input energy is above ~several hundred kJ to 1 MJ for the fuel compression of the reactor size target. Therefore, it is pointed out that the larger input energy has an advantage to release the fusion energy in a fusion reactor system.

8. Discussions

In the above sections, the important issues are studied to obtain the sufficient target gain. In addition to above results, there are other points to be considered, for example, the total fuel mass and the R-T instability. If a target contains a too-much fuel for the fixed input energy, the fuel cannot be accelerated sufficiently, the implosion efficiency becomes lower and the ignition may not to be attained. Therefore, it is important to optimize the DT fuel mass for the high gain[13].

Another important issue is the uniform implosion. The R-T instability prevents the target uniform implosion and gives the upper limit for the target radius. The non-uniform beam illumination and the target non-uniformity itself also introduce the target non-uniform implosion. The numerical analyses for the effect of the non-uniform implosion on $\rho R$ have been done by the 3-D numerical computer code[11]. The previous results[11] show that the restrictions for the non-uniformity of the implosion acceleration should be less than 2.7% for the volume compression ratio of 10000 or ~3 - 6% for 1000, as shown in Fig. 12.

Acknowledgements

The authors would like to present their appreciations to Prof. Keishiro Niu, who has passed away on April 17, 2012, for his strong encouragements for ion beam nuclear fusion. Prof. K. Niu has started the ion beam fusion since 1978 together with the authors in Tokyo Inst. of Tech., and has continued to encourage them to explore the ion beam inertial fusion.

References

1) K. Kasuya, K. Horioka, T. Takahashi, A. Urai and M. Hijikawa, Appl. Phys. Lett., 39 (1981)887.
2) S. Miyamoto, A. Yoshinouchi, et. al., Jpn. J. Appl. Phys. 22 (1983) L703.
3) P. A. Miller, et. al., Laser and Particle Beams, 2 (1984) 153.
4) S. Kawata, K. Niu and H. Murakami, Jpn. J. Appl. Phys., 22 (1983) 302.

5) J. R. Freeman, L. Baker and D. L. Cook, Nucl. Fusion, 22 (1982) 383.
6) H. Murakami, S. Kawata and K. Niu, Jpn. J. Appl. Phys., 22 (1983) 305.
7) D. G. Colombant, S. A. Goldstein and D. Mosher, Phys. Rev. Lett., 45 (1980) 1253.
8) S. Kawata and K. Niu, Laser and Particle Beams, 1 (1983) 121.
9) D. G. Colombant, S. A. Goldstein, NRL Meorandum Report 4640 (1981).
10) M. Tamba, N. Narata, S. Kawata and K. Niu, Laser and Particle Beams, 1 (1983) 121.
11) S. Kawata and K. Niu, J. Phys. Soc. Jpn., 53 (1984) 3416.
12) S. Ichimaru, *Basic Principles of Plasma Physics*, (W. A. Benjamine, 1973) P.64
13) S. Kawata and K. Niu, Res. Rep. in Inst. of Plasma Phy., Nagoya Univ., IPPJ-502 (1981).

Table 1 Parameter values employed in the numerical computations as the typical values. The notation $C_{AL}$ shows the ratio of the deposited input energy in the A1 layer to the total input energy $E_b$. The input beam power varies with $(time)^{2.5}$ from zero. The same time dependence of the IB input power is employed for the numerical computations through the paper.

| | | |
|---|---|---|
| INPUT BEAM ENERGY | $E_b$ | 5 MJ |
| BEAM-PROTON ENERGY | $e_b$ | 5 MeV |
| BEAM DURATION TIME | $\tau_b$ | 35 nsec |
| MAXMAUM PROTON INCIDENT ANGLE | $e_m$ | 60 degrees |
| BEAM POWER | | $t^{2.5}$ |
| TARGET RADIUS | $r_t$ | 5 mm |
| | $C_{AL}$ | 0.8 |
| TOTAL DT MASS | $M_{DT}$ | 1.0 mg |

Table 2  Parameter values employed in the computations to check the effect of the IB oblique incidence to the target surface on the implosion performance.

| | | |
|---|---|---|
| INPUT BEAM ENERGY | $E_b$ | 7 MJ |
| BEAM-PARTICLE ENERGY | $e_b$ | 5 MeV |
| BEAM DURATION TIME | $\tau_b$ | 35 nsec |
| BEAM POWER | | $t^{2.5}$ |
| TARGET RADIUS | $r_t$ | 5 mm |
| | $C_{Al}$ | 0.8 |
| TOTAL DT MASS | $M_{DT}$ | 4.0 mg |

Table 3  Numerical results. The parameter values employed are listed in the Table, and the others, which are not listed, are the same as those in Table 1.

PLASMA EFFECT ON BEAM STOPPING POWER

| Target radius | $r_t$ | = | 5 mm |
|---|---|---|---|
| Particle energy | $e_b$ | = | 5 MeV |
| Beam energy | $E_b$ | = | 7 MJ |
|  | $C_{Al}$ | = | 0.8 |
| Beam duration time | $\tau_b$ | = | 35 nsec |

|  |  | $\theta_m$ = 0.0  $M_{DT}$ = 4 mg | | $\theta_m$ = 60.0  $M_{DT}$ = 2 mg | |
|---|---|---|---|---|---|
| PLASMA EFFECT |  | ON | OFF | ON | OFF |
| $\eta_{KDT}$ | (%) | 1.92 | 3.04 | 0.72 | 1.27 |
| $\eta_{Kin}$ | (%) | 10.6 | 7.0 | 10.9 | 9.07 |
| $\eta_{imp}$ | (%) | 7.4 | 4.62 | 4.33 | 4.17 |
| $M_{Alin}$ at $\tau_b$ | (mg) | 72.7 | 54.3 | 83.8 | 72.7 |
| $MV_{Alin} + MV_{DT}$ at $\tau_V$ | (g cm/s) | $-7.93 \times 10^5$ | $-5.85 \times 10^5$ | $-7.16 \times 10^5$ | $-5.39 \times 10^5$ |
| $<V_{DT}>/10^7$ | (cm/s) | 2.59 | 3.25 | 2.25 | 2.98 |
| $<V_{in}>/10^7$ | (cm/s) | 1.03 | 1.02 | 0.834 | 0.722 |
| $\tau_V$ | (nsec) | 46.6 | 37.0 | 51.1 | 43.4 |
| $<T_{Al}>$ at $\tau_V$ | (eV) | 205 | 364 | 155 | 259 |
| $E_{TAl}$ at $\tau_V$ | (MJ) | 2.41 | 4.30 | 1.82 | 3.06 |
| Q |  | 120 | $2.70 \times 10^{-4}$ | 77.3 | 68.5 |
| $\rho R$ | (g/cm$^2$) | 13.8 | 7.16 | 22.0 | 13.7 |

Table 4. Numerical results. The computations are performed to check the effect of the difference of the particle energy $e_b$. $E_{TAl}$ shows the retained thermal energy in the Al layer at the void closure time and $E_{TPb}$ the thermal energy retained in the Pb Layer. $E_T = E_{TAl} + E_{TPb}$.

PARTICLE ENERGY ($e_b$) EFFECT

| Target radius | $r_t$ | = | 5 mm |
| Beam input energy | $E_b$ | = | 5 MJ |
| Beam duration time | $\tau_b$ | = | 35.0 nsec |
| Beam incident angle | $\theta_m$ | = | 60.0 |
| Total DT mass | $M_{DT}$ | = | 1 mg |
| | $C_{Al}$ | = | 0.8 |

| Proton Particle energy $e_b$ | | 5 MeV | 3 MeV |
|---|---|---|---|
| $\eta_{KDT}$ | (%) | 0.44 | 0.794 |
| $\eta_{imp}$ | (%) | 4.92 | 3.23 |
| $\eta_{Kin}$ | (%) | 11.5 | 9.95 |
| $<V_{DT}>/10^7$ | (cm/s) | 2.1 | 2.82 |
| $M_{Alin}$ | (mg) | 87.6 | 30.0 |
| $M_{Al}$ | (mg) | 110 | 47.5 |
| $M_{Pb}$ | (mg) | 106 | 50.8 |
| $\dfrac{M_{Pb} + (M_{AL} - M_{ALin}) \times 100}{M_{Pb} + M_{AL} + M_{DT}}$ % | | 59.2 | 68.8 |
| $E_{TAl}$ at $\tau_v$ | (MJ) | 1.32 | 1.14 |
| $E_{TPb}$ at $\tau_v$ | (MJ) | 1.61 | 1.81 |
| $E_T$ at $\tau_v$ | (MJ) | 2.49 | 2.72 |
| $<T_{DT}>$ at $\tau_v$ | (eV) | 4.07 | 25.5 |
| $<T_{Al}>$ at $\tau_v$ | (ev) | 112 | 219 |
| $<T_{Pb}>$ at $\tau_v$ | (ev) | 174 | 342 |
| $\tau_v$ | (nsec) | 55.5 | 46.6 |
| Q | | 59 | 49 |
| $\rho R$ | (g/cm$^2$) | 22.4 | 12.2 |

Table 5. Numerical results. The computations are carried out to investigate the effect of the difference of the LIB duration time on the implosion performance.

Beam DURATION ($\tau_b$) EFFECT

| Target radius | $r_t$ | = | 5 mm |
| | $C_{Al}$ | = | 0.8 |
| Total DT mass | $M_{DT}$ | = | 1 mg |
| Beam input energy | $E_b$ | = | 5 MJ |
| Beam particle energy | $e_b$ | = | 5 MeV |

| Beam duration time $\tau_b$ (nsec) | | 35 | 55 |
|---|---|---|---|
| $\eta_{KDT}$ | (%) | 0.44 | 0.349 |
| $\eta_{imp}$ | (%) | 4.5 | 3.67 |
| $\eta_{Kin}$ | (%) | 11.4 | 7.85 |
| $<T_{DT}>$ at $\tau_v$ | (ev) | 4.07 | 2.86 |
| $<V_{DT}>/10^7$ at $\tau_v$ | (cm/s) | 2.09 | 1.87 |
| $\tau_v$ | (nsec) | 55.5 | 72.1 |
| $E_{TAL}$ at $\tau_v$ | (MJ) | 1.33 | 1.54 |
| $\rho R$ | (g/cm$^2$) | 22.4 | 29.3 |
| Q | | 59 | 59.5 |

Table 6. Numerical results. The computations are carried out to check the effect of the difference of the target radius on the implosion performance.

TARGET RADIUS ($r_t$) EFFECT

| Total DT mass | $M_{DT}$ | = | 1 mg |
|---|---|---|---|
| | $C_{Al}$ | = | 0.8 |
| Beam input enrgy | $E_b$ | = | 5 MJ |
| Beam particle enrgy | $e_b$ | = | 5 MeV |
| Beam duration time | $\tau_b$ | = | 35 nsec |
| Beam incident angle | $\theta_m$ | = | 60 |

| Target radius $r_t$ | (cm) | 0.5 | 0.35 |
|---|---|---|---|
| $\eta_{KDT}$ | (%) | 0.44 | 0.642 |
| $\eta_{imp}$ | (%) | 4.5 | 4.34 |
| $\eta_{Kin}$ | (%) | 11.4 | 6.8 |
| $<V_{DT}>$ $10^7$ at $\tau_v$ | (cm/s) | 2.09 | 2.53 |
| $E_{TAl}$ at $\tau_v$ | (MJ) | 1.33 | 1.42 |
| $\tau_v$ | (nsec) | 55.5 | 40.7 |
| $M_{Al}$ | (mg) | 110 | 53 |
| $\rho R$ | (g/cm$^2$) | 22.4 | 15.1 |
| Q | | 59 | 51.2 |

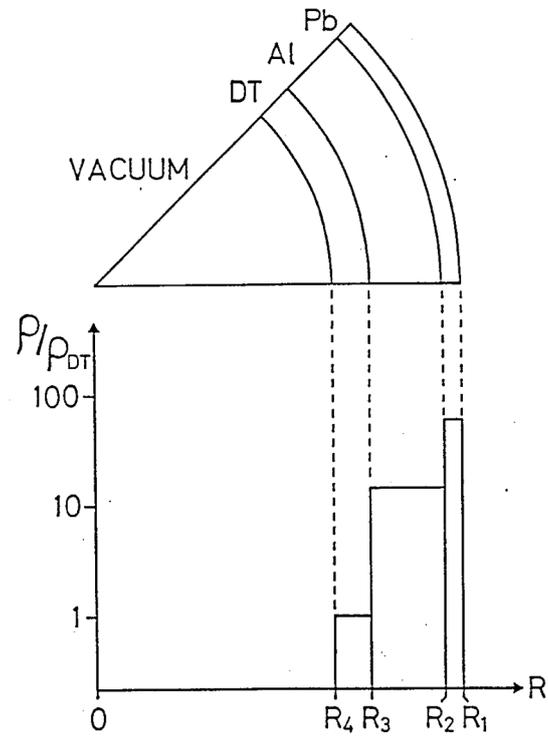

Fig. 1 The IB-target structure. The input IB deposits its energy mainly into the Al energy absorber layer.

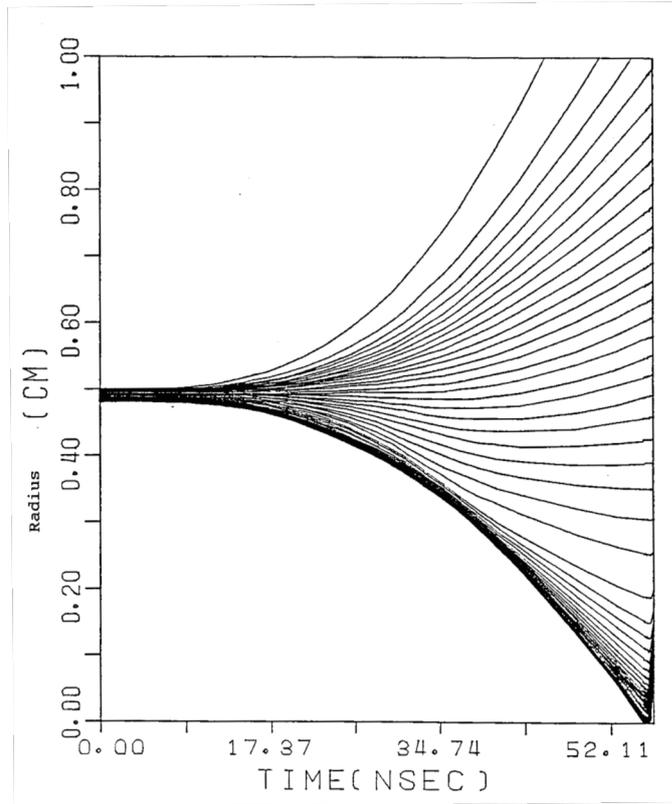

Fig. 2  The stream lines, obtained by the numerical computations in the case of the parameter values shown in Table 1. After the void closure time, the DT fuel is compressed further more by the Al pusher.

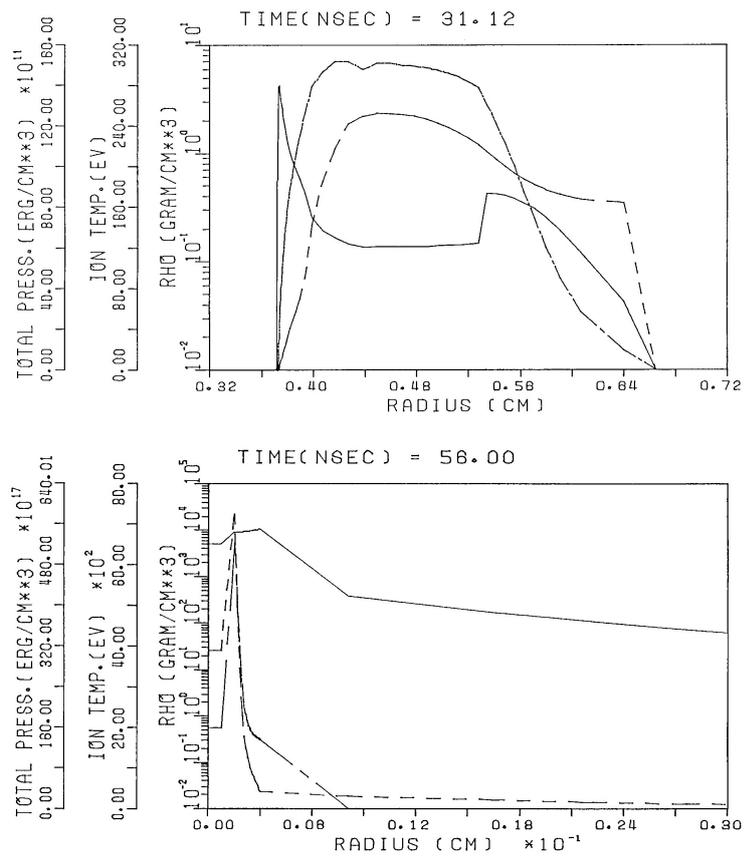

Fig. 3 The space profiles of the ion temperature $T_i$, the mass density $\rho$ and the pressure $P$.

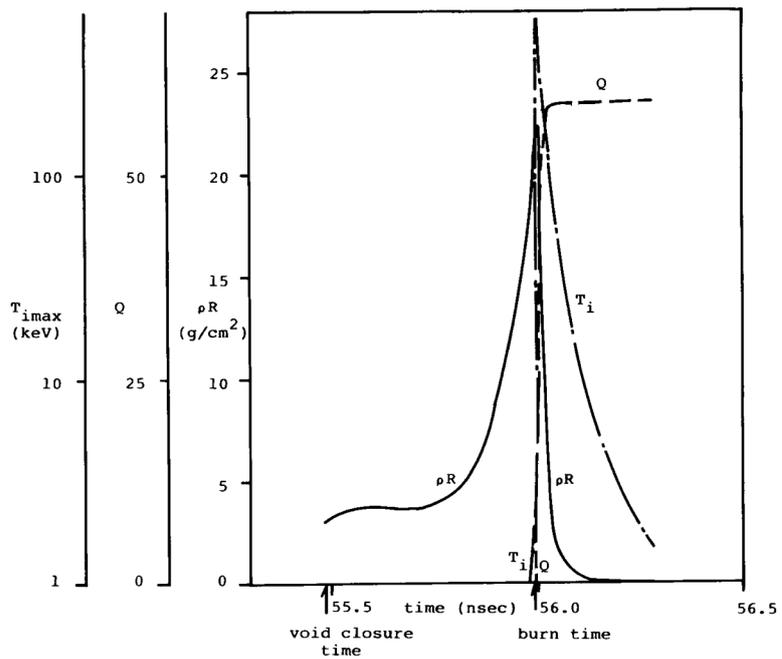

Fig. 4 The numerical results of the time sequences of the DT ion peak temperature, the pellet gain Q and the fuel density-radius product ρR.

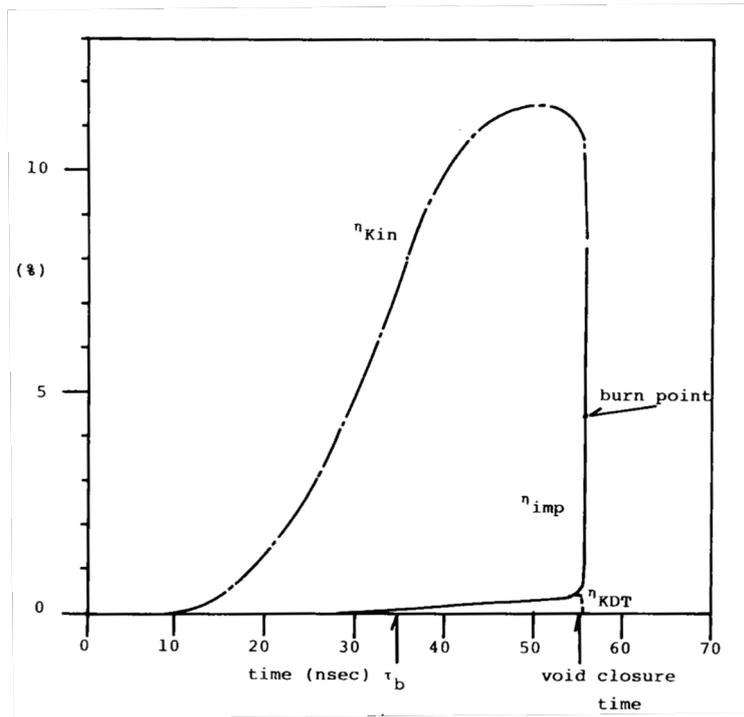

Fig. 5 Time histories of the implosion efficiency $\eta_{imp}$, the DT kinetic energy ratio $\eta_{KDT}$ and ratio of the inward-moving kinetic energy $\eta_{Kin}$.

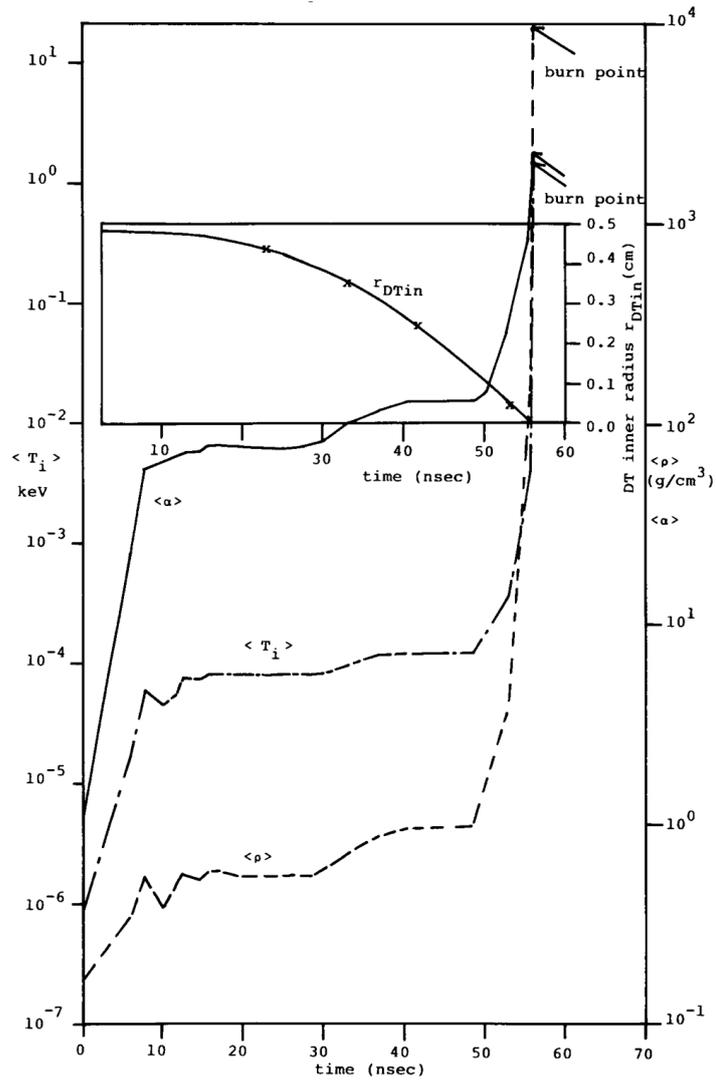

Fig. 6 DT fuel preheating feature. The notation $<\alpha>$ is defined by (the spatially averaged ion temperature of DT fuel at time $t$) / (the spatially averaged ion temperature estimated by the adiabatic relation for the initial pressure and the density at $t$).

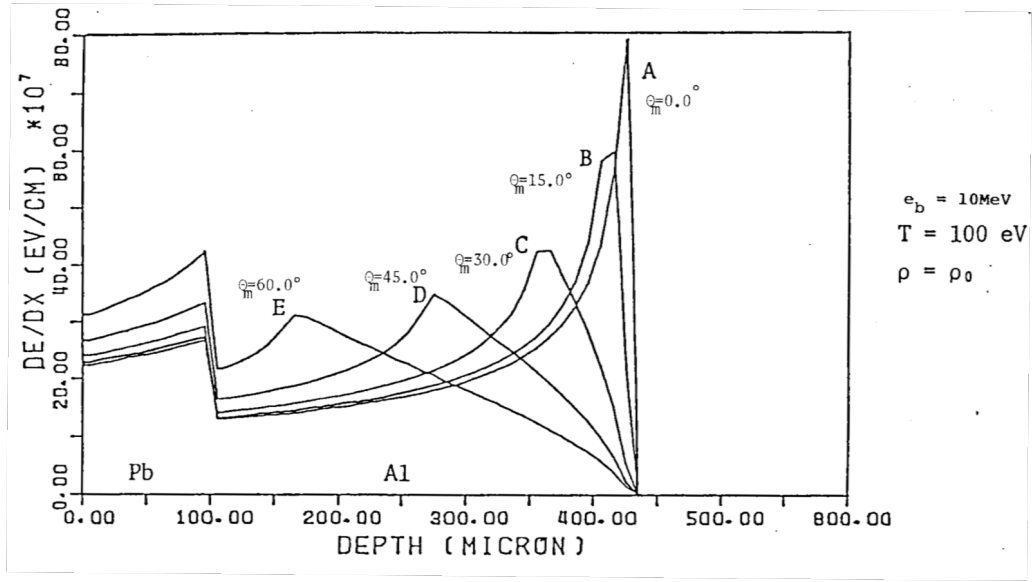

Fig. 7 The space profiles for the IB (ion beam) energy deposition. The profiles are computed by the following assumptions: the IB has the homogeneous distribution of the incident angle to the target surface normal in the azimuthal and polar directions, and the profiles are averaged in the both directions. In the figure, the target density is the solid one, the ion and electron temperatures 100eV and the input proton energy 10 MeV.

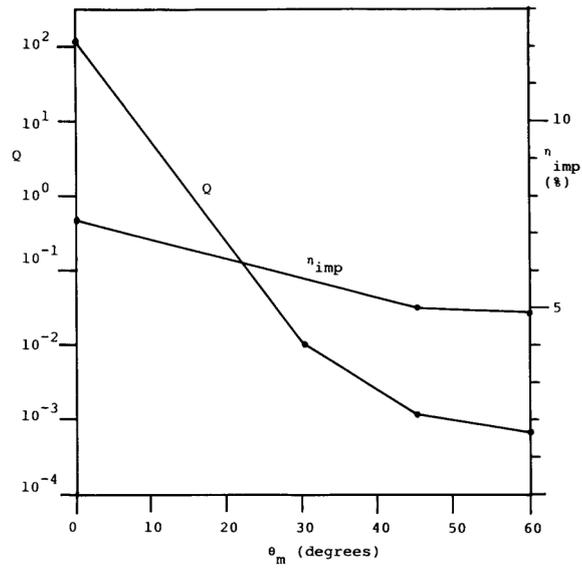

Fig. 8 The relation between the implosion efficiency $\eta_{\text{imp}}$ and the incident angle $\theta_{\text{m}}$. The pellet gain $Q$ is also presented.

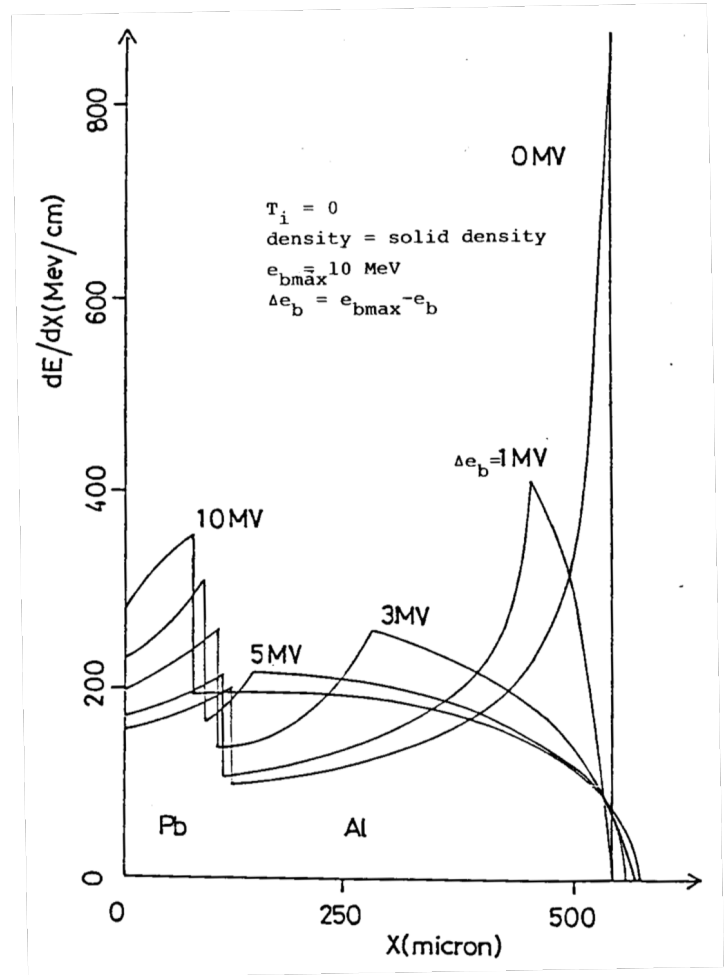

Fig. 9 The space profiles of the IB deposition energy. The input protons have the broad and homogeneous energy spectrum from $e_b$ to $e_{bmax}(=10$ MeV). The $\Delta e_b$ is defined by $\Delta e_b = e_{bmax} - e_b$.

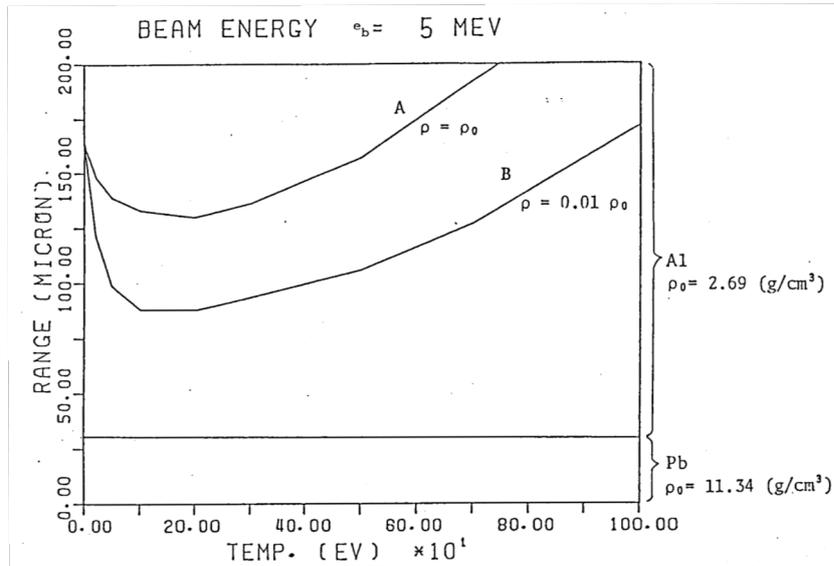

Fig. 10 The stopping range versus the target temperature. The stopping range shortening is clearly presented. The range shortening comes from the plasma effect of the stopping power.

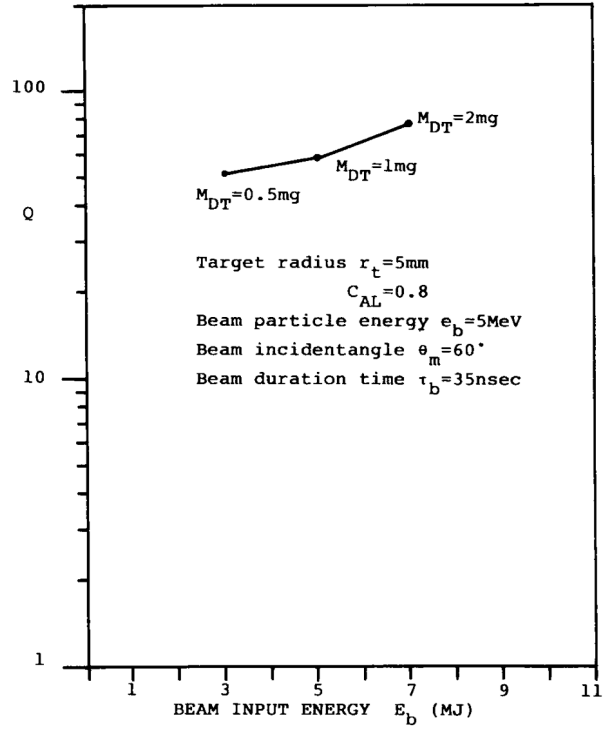

Fig. 11 The target gain $Q$ versus the IB input energy $E_b$. In the figure the results are optimized only for the DT fuel total mass $M_{DT}$.

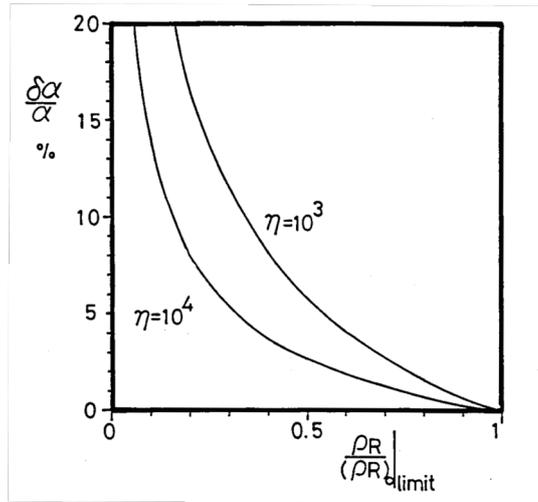

Fig. 12 The non-uniformity effect on the target implosion. The relation between the non-uniformity of the implosion acceleration $\delta\alpha$ and mass-density radius product $\rho R$. In the figure $\eta$ shows the volume compression ration (usually $\eta \sim 10^3 \sim 10^4$).